\documentclass[showpacs,prl,aps,twocolumn]{revtex4-1}

\pdfoutput=1

\usepackage{amsmath,amsfonts,amssymb,bm,graphicx,hyperref}

\newcommand{\vt}{\vartriangle}

\newcommand{\ts}{\tilde{\sigma}}

\newcommand{\s}{\tilde{s}}
\newcommand{\tK}{\tilde{K}}

\begin{document}

\title{Transition in coupled replicas may not imply a finite temperature ideal glass transition in glass forming systems}

\author{Juan P. Garrahan}
\affiliation{School of Physics and Astronomy, University of Nottingham, Nottingham, NG7 2RD, United Kingdom}
\date{\today}

\begin{abstract}
A key open question in the glass transition field is whether a finite temperature thermodynamic transition to the glass state exists or not.  Recent simulations of coupled replicas in atomistic models have found signatures of a static transition as a function of replica coupling.  This can be viewed as evidence of an associated thermodynamic glass transition in the uncoupled system.  We demonstrate here that a different interpretation is possible.  We consider the triangular plaquette model, an interacting spin system which displays (East model-like) glassy dynamics in the absence of any static transition.  We show that when two replicas are coupled there is a curve of equilibrium phase transitions, between phases of small and large overlap, in the temperature-coupling plane (located on the self-dual line of an exact temperature/coupling duality of the system) which ends at a critical point.  Crucially, in the limit of vanishing coupling the finite temperature transition disappears, and the uncoupled system is in the disordered phase at all temperatures.  We discuss interpretation of atomistic simulations in light of this result.
\end{abstract}

\maketitle

A series of recent simulation works \cite{Cammarota2010,Berthier2013,Parisi2013} have studied in some detail the thermodynamic properties of (reasonably realistic) glass-forming systems when two copies (or replicas) of the system are {\em coupled}.  These studies have produced evidence for an equilibrium transition between static phases of small and large overlap between the replicas \cite{Cammarota2010,Berthier2013,Parisi2013}.  This phase transition seems to be first-order and the coexistence curve in the temperature (or density, in the case of hard-spheres) and replica coupling plane terminates at a critical point  \cite{Berthier2013,Parisi2013}.  Such a behaviour of the coupled system is consistent with theoretical predictions \cite{Franz1997,Franz1998,Biroli2013b,Franz2013} from the random first-order transition theory (RFOT) \cite{Mezard2000,Lubchenko2007,Parisi2010}.  Furthermore, these numerical results could be interpreted as evidence for the existence of a thermodynamic transition to the glass state in the {\em uncoupled} system.  This is an important issue, since a central unresolved question
in the understanding of the kinetic phenomenon we term ``glass transition''  \cite{Ediger1996,Cavagna2009,Berthier2011,Biroli2013} is whether it is caused by an underlying thermodynamic singularity or not.  In particular, it may seem that the observations of \cite{Berthier2013,Parisi2013} are difficult to reconcile with purely dynamical approaches such as dynamic facilitation \cite{Chandler2010}.  The purpose of this paper is to show that this is not the case, and that it is possible to explain the behaviour of coupled replicas without recourse to a thermodynamic singularity in the uncoupled system.

\begin{figure}[t!]
\includegraphics[width=0.8\columnwidth]{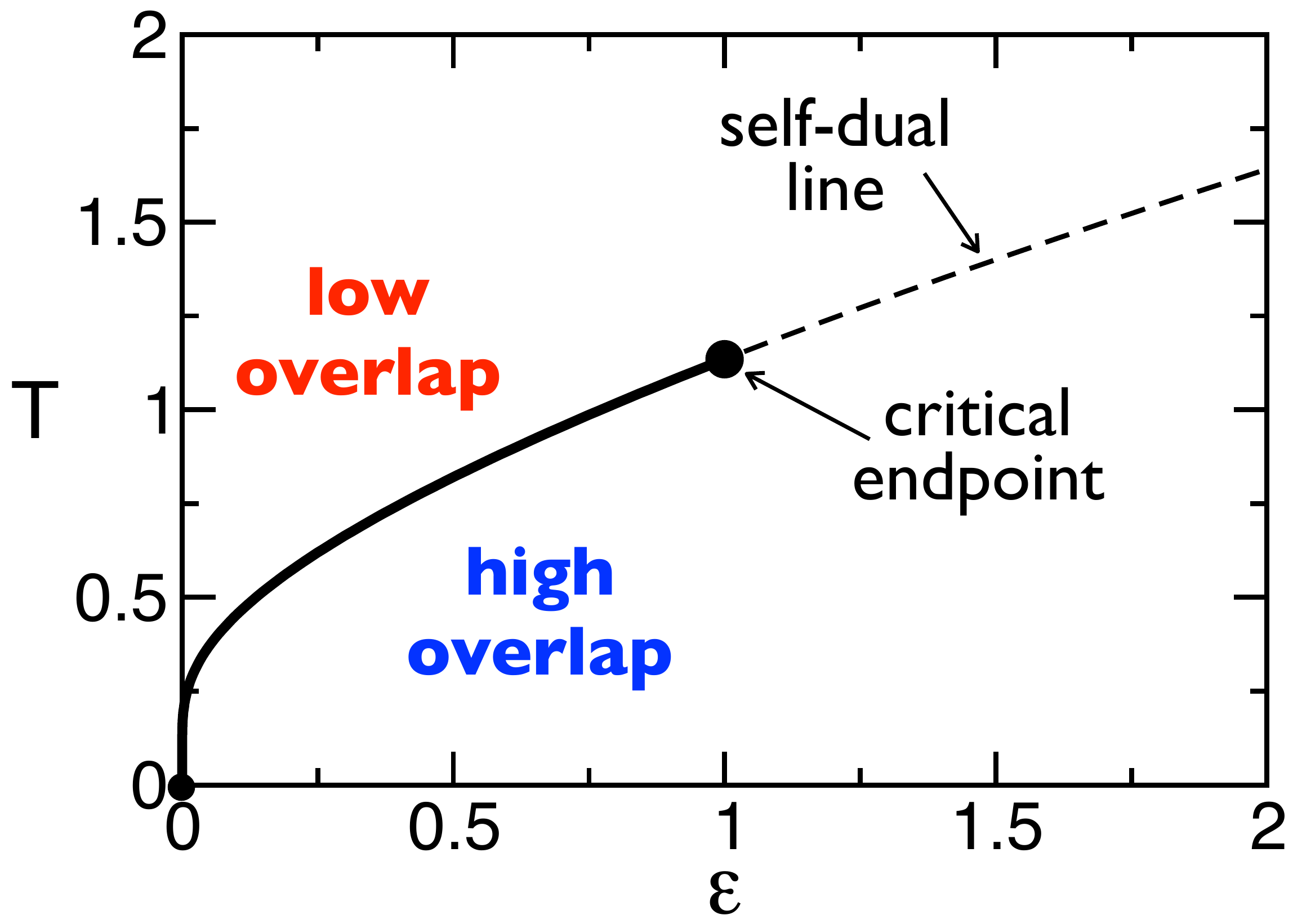}
\caption{Exact phase diagram of two coupled triangular plaquette glass models.  The temperature, $T$, and coupling, $\epsilon$, are in units of the energy constant of the model, $J$.  The full line is a curve of first-order transitions from a phase with low overlap between the replicas to a phase of high overlap between the replicas.  The transition line, $\bar{T}(\varepsilon)$, is on the self-dual curve of the model (dashed line), and ends at a critical point $(\varepsilon_{\rm c}/J,T_{\rm c}/J)=[1,1/\log{(1+\sqrt{2})}]$.  The phase transition disappears in the limit of vanishing coupling, $\bar{T}(\varepsilon \to 0)=0$ and there is no finite $T$ singularity in the uncoupled system.
}
\label{fig1}
\end{figure}

Our results are summarised in Fig.\ 1.  The figure shows the phase diagram in the temperature, $T$, and replica coupling, $\varepsilon$, plane for two coupled copies of a specific glass model, the triangular plaquette model (TPM) \cite{Newman1999,Garrahan2000,Garrahan2002,Ritort2003,Sasa2010}.  The TPM is one of a class of interacting spin models which map to kinetically constrained models \cite{Garrahan2000,Garrahan2002,Ritort2003}, i.e.\ that display complex glassy dynamics with simple thermodynamics.  The dynamics of the TPM in particular is that of the East model \cite{Jackle1991,Garrahan2000,Garrahan2002,Ritort2003}, the facilitated model which captures best the phenomenology of glass formers \cite{Elmatad2009,Keys2013}.  Figure 1 shows that the TPM also has a first-order transition between static phases of low and high overlap occuring at a temperature $\bar{T}$ that depends on the coupling, $\bar{T}=\bar{T}(\varepsilon)$.  The curve $\bar{T}(\varepsilon)$ terminates at a critical point $(\varepsilon_{\rm c}/J,T_{\rm c}/J)$.  Crucially, the coexistence curve meets the $T$ axis at $T=0$, that is, $\bar{T}(\varepsilon \to 0)=0$, and the transition vanishes in the absence of coupling.  The upshot is two-fold: (i) from a facilitation perspective it is indeed possible to account for the static behaviour observed in simulations at finite coupling $\varepsilon$; (ii) the static transition may disappear when the coupling vanishes, and thus one needs to be carful about extrapolations of finite $\varepsilon$ results for interpretations of what occurs in the uncoupled system.

The TPM is defined by the energy function \cite{Newman1999,Garrahan2000,Garrahan2002,Ritort2003}
\begin{equation}
E(s) \equiv -\frac{J}{2} \sum_{\bigtriangleup} s_{i} s_{j} s_{k} ,
\label{E}
\end{equation}
where $s_{i}=\pm 1$ is a classical Ising spin on the site $i$ of a triangular lattice ($i = 1,\ldots,N$), and where the sum is over upward pointing triangles $\vartriangle$, or plaquettes, of the lattice.  For simplicity we consider periodic boundary conditions, at least in one direction of the lattice, and such that the linear size in that direction is a power of two, $L=2^{K}$.  In this case the ground state of (\ref{E}) is unique \cite{Newman1999,Garrahan2000,Garrahan2002,Ritort2003}, corresponding to all spins pointing up.  (Other boundary conditions lead to the same physics, but the mathematics is more tedious \cite{Garrahan2002}.) One can define plaquette spin variables, $\tau_{\vartriangle} \equiv - s_{i} s_{j} s_{k}$, for all plaquettes $\vartriangle$ of the lattice.  In terms of these variables the ground state is $\tau_{\vartriangle} = -1$ for all $\vartriangle$, and the elementary excitation of the system corresponds to a defective plaquette, $\tau_{\vartriangle} = +1$.  For the chosen boundary conditions, there is a one-to-one mapping between spins $s_{i}$ and plaquettes $\tau_{\vartriangle}$, and the thermodynamics is that of a free gas of binary plaquettes \cite{Newman1999,Garrahan2000,Garrahan2002,Ritort2003}.  While the thermodynamics of the TPM is trivial its dynamics is not.  This is because an elementary dynamical move corresponds to a spin-flip, $s_{i} \to - s_{i}$, and each spin-flip reverts the three plaquettes $\tau_{\vartriangle}$ in which it participates.  The plaquette dynamics is kinetically constrained \cite{Garrahan2000,Garrahan2002,Ritort2003} leading to complex glassy dynamics at low temperatures.  In the case of the TPM the effective dynamics is that of the East facilitated model \cite{Garrahan2000,Garrahan2002,Ritort2003}, and therefore displays ``parabolic'' super-Arrhenius growth of timescales, temperature dependent stretching of correlations, dynamic heterogeneity, and other hallmarks of the glass transition \cite{Biroli2013}. 

We now consider two TPMs, with a coupling between them that is proportional to their overlap.  The total energy of the coupled system is defined as,
\begin{equation}
E(s^{a},s^{b}) \equiv -\frac{J}{2} \sum_{\bigtriangleup} \left( s_{i}^{a} s_{j}^{a} s_{k}^{a} + s_{i}^{b} s_{j}^{b} s_{k}^{b} \right) - \varepsilon \sum_{i} s_{i}^{a} s_{i}^{b},
\label{E2}
\end{equation}
where $s^{a}$ and $s^{b}$ are the spin configurations in the replicas $a$ and $b$.  The coupling field $\varepsilon$  is conjugate to the overlap, $q(s^{a},s^{b}) \equiv N^{-1} \sum_{i} s_{i}^{a} s_{i}^{b}$, which measures the similarity of the spin configurations of the two copies.  Equation (\ref{E2}) corresponds to the {\em annealed} coupling, in the sense that both replicas are considered on equal footing.  The {\em quenched} coupling, in contrast, corresponds to when one of the replicas is frozen in an equilibrium configuration.  We restrict here to the annealed case which is easier to treat analytically.  Just like in previous works \cite{Cammarota2010,Berthier2013,Parisi2013}, the assumption is that the behaviour of the simpler annealed problem is a good qualitative indicator of what happens in the more complex quenched problem.  

The partition sum for the coupled replicas reads,
\begin{equation}
Z(K_{1},K_{2}) \equiv \sum_{\{s^{a},s^{b}\}} e^{K_{1} \sum_{\bigtriangleup} \left( s_{i}^{a} s_{j}^{a} s_{k}^{a} + s_{i}^{b} s_{j}^{b} s_{k}^{b} \right) + K_{2} \sum_{i} s_{i}^{a} s_{i}^{b}} ,
\nonumber
\end{equation}
with $K_{1} \equiv J/2 k_{\rm B} T$ and $K_{2} \equiv \varepsilon/k_{\rm B} T$.  By applying techniques similar to those of Refs.\ \cite{Sasa2010,Deng2010} it is easy to prove the existence of an exact duality that relates the partition sum at state point $(K_{1},K_{2})$ to that at $(K_{1}^{*},K_{2}^{*})$  (see Methods for details).  That is,
\begin{equation}
Z(K_{1},K_{2}) = \left( \sinh{2K_{1}} \sinh{K_{2}} \right)^{N} Z(K_{1}^{*},K_{2}^{*}),
\label{Z}
\end{equation}
where the transformation that connects the two state points is given by
\begin{equation}
e^{-K_{2}^{*}} = \tanh{K_{1}}, \;\;\; \tanh{K_{1}^{*}} = e^{-K_{2}} .
\label{K}
\end{equation}
The duality relation (\ref{Z})-(\ref{K}) for the problem of two coupled TPMs is analogous to that of the TPM in a magnetic field \cite{Sasa2010}, which in turn is the one of the generalized Baxter-Wu model \cite{Deng2010}.  

The {\em self-dual} line is defined by the condition that the partition functions $Z(K_{1},K_{2})$ and $Z(K_{1}^{*},K_{2}^{*})$ coincide.  From (\ref{Z}) we see that this happens when,
\begin{equation}
\left( \sinh{\frac{J}{T}} \right) \left( \sinh{\frac{\varepsilon}{T}} \right) = 1,
\label{ss}
\end{equation}
where we have reverted to the original parameters of the problem and set $k_{\rm B}$ to unity.  Figure 1 shows the self-dual line in the $(\varepsilon,T)$ plane (for unit $J$, which just sets the energy scale of the problem).  Note that this line approaches the origin as $\varepsilon \sim 2 T e^{-J/T}$.  

The self-dual line (\ref{ss}) is the locus of first-order phase transitions, just like it happens in the generalised Baxter-Wu model \cite{Deng2010} or the TPM in a field \cite{Sasa2010}.  These occur at a coupling dependent temperature $\bar{T}(\varepsilon)$, which obeys Eq.\ (\ref{ss}).  For a given $\varepsilon$, the transition at $\bar{T}(\varepsilon)$ is between a phase with low overlap between the replicas, at $T > \bar{T}(\varepsilon)$, to one of high overlap, at $T < \bar{T}(\varepsilon)$.  The transition line ends at a critical point $(\varepsilon_{\rm c},T_{\rm c})$.  

The location of the critical point can be obtained from the following argument.  The duality transformation (\ref{K}) relates a state point $(\varepsilon/J,T/J)$ to another state point $[(\varepsilon/J)^{*},(T/J)^{*}]$, while never connecting the half-planes $\varepsilon > J$ and $\varepsilon < J$.  That is, if $\varepsilon/J > 1$ (resp.\ $<1$) then $(\varepsilon/J)^{*} > 1$ (resp.\ $<1$).  The reason is that $\varepsilon=J$ is a special point: for this value of the coupling the problem defined by Eq.\ (\ref{E2}) becomes equivalent to the Baxter-Wu model \cite{Baxter2007}.  This model has a critical transition at some temperature $T_{c}$, which we can obtain from the self-dual condition (\ref{K}) when $\varepsilon_{c} = J$, 
\begin{equation}
\left( \sinh{\frac{J}{T_{c}}} \right)^{2} = 1 
\; \Rightarrow \; (\varepsilon_{\rm c},T_{\rm c}) = \left(J,\frac{J}{\log{(1+\sqrt{2})}}\right) .
\label{Tc}
\end{equation}
The first-order line of (\ref{E2})-(\ref{Z}) will end at this critical point, see Fig.\ 1.  Note that while the critical temperature of the Baxter-Wu model is the same as that of the two-dimensional Ising model, its critical properties are those of the two-dimensional four-state Potts model \cite{Baxter2007}. 

The structure of the phase diagram of Fig.\ 1 is interesting as it offers an alternative interpretation on the recent numerical results on replicated systems \cite{Cammarota2010,Berthier2013,Parisi2013}.  The supercooled regime of the TPM is that of $T/J < 1$.  Within this regime, Fig.\ 1 shows that the coexistence coupling $\bar{\varepsilon}$ changes rapidly with temperature.  If one were to simulate the coupled TPMs for a range of temperatures satisfying $1 \gtrsim T/J \gg 0$ (over which the relaxation time of the uncoupled TPM would still grow by many orders of magnitude), it would be easy to conclude, incorrectly, that the coexistence line would extrapolate to cross the $T$ axis at some finite temperature.

The results presented here highlight some of the differences between theoretical approaches to the glass transition \cite{Berthier2011,Biroli2013}, and in particular RFOT \cite{Mezard2000,Lubchenko2007,Parisi2010} versus dynamical facilitation \cite{Chandler2010}.  In order to have a thermodynamic transition to an ideal glass state at non-zero temperature, as RFOT advocates \cite{Mezard2000,Lubchenko2007,Parisi2010}, it is necessary to have excitations to low energy states which are extended \cite{Bouchaud2004} rather than pointlike. Extended excitations would be needed, on one hand to support interfaces, and on the other to prevent their proliferation due to entropy, which would destroy the transition.  

In contrast, the central tenet of the facilitation approach is that excitations are localised \cite{Chandler2010,Keys2011}.  The entropy of mixing of such excitations destabilises any ordered (or randomly frozen) state \cite{Stillinger1988}.  Plaquette models like the TPM are a clear example of this mechanism. The plaquette interactions of the energy function (\ref{E}) can be minimised by a multiplicity of local spin arrangements, four per plaquette in the case of the TPM (while there is one, or few depending on boundaries \cite{Garrahan2002}, global energy minima).  Whenever there is a mismatch between such local arrangements the energy cost is localised in an excited plaquette.  Defects are thus pointlike rather than extended.  The ``deconfinement'' of defects is always entropically favoured, and the TPM and similar plaquette models are disordered at all temperatures \cite{Newman1999,Garrahan2000,Garrahan2002,Ritort2003}.  Structural rearrangement is concentrated around defects \cite{Garrahan2000,Garrahan2002} and these systems show all the features of facilitated dynamics \cite{Garrahan2002b,Chandler2010}.  
(These mechanisms are not different from those of other systems with local constraints, such as dimer coverings \cite{Garrahan2009} or spin-ice \cite{Mostame2013}.)

The introduction of a magnetic field \cite{Sasa2010}, or other interactions \cite{Jack2005}, can lead to a confinement of defects which then allows for a transition to an ordered state.  This is essentially what happens for two coupled TPMs: the  
coupling between replicas can induce a confinement of the difference between the defect arrangements in the two replicas, leading not to an ordered state in either replica, as in the single TPM in a field, but to a phase of large overlap between the two copies.    
A finite coupling, however, is required at all non-zero temperatures to keep this confinement and prevent the two copies from becoming independent. This contrasts to what RFOT would predict occurs for glass formers.  It is also noteworthy, that the TPM is one of several simple systems, with local and disorder-free interactions (the East model is another example), which can be solved in finite dimensions and be shown to display the ideal behaviour of the dynamical facilitation approach. 
This is a further contrast with RFOT where solvable low-dimensional models that furnish that perspective are yet to be found \cite{Yeo2012,Cammarota2012}. 

In summary, we have shown that when two replicas of a triangular plaquette model are coupled there can be a first-order transition, at finite coupling strength, between a phase of low overlap where the two copies are idependent, to one of high overlap where the two copies are correlated.  At high enough temperature this transition ends at a critical point.  These are the qualitative features revealed by numerical simulations of replicated glass forming models \cite{Cammarota2010,Berthier2013,Parisi2013}.  A key observation here is that when the coupling is removed the phase transition disappears.  There are two messages to take from these results: first, the numerically observed behaviour of replicated liquids is not inconsistent with dynamic facilitation, as an ideal model from this approach displays essentially the same behaviour; and second, it may not be safe to assume that a transition at finite coupling implies a finite temperature/density thermodynamic glass transition in the uncoupled system.

\noindent
{\bf Methods.}  Here we prove the duality relations quoted in the text above.  We follow closely the notation of Ref.\ \cite{Sasa2010}.  The spins $s_i$ live on the sites $i \in \Lambda$ of the triangular lattice $\Lambda$.  The plaquettes in turn are located on the sites $\vt \in \Lambda^*$ of the dual lattice $\Lambda^*$.  The plaquette variable $\tau_\vt$ is a function of the spins immediately to the left, right and top of $\vt$.  If we denote these sites in $\Lambda$ by $l_\vt,r_\vt,t_\vt$, respectively, then we have $\tau_\vt = - s_{l_\vt} s_{r_\vt} s_{t_\vt}$.   Similarly, the spin $s_i$ participates in three plaquettes immediately to the left, right and bottom of $i$.  We denote these sites on $\Lambda^*$ by $l_i,r_i,b_i$, respectively.

Just like in Refs.\ \cite{Deng2010,Sasa2010} for the cases of the generalised Baxter-Wu model or the TPM in a field, duality is proved by introducing extra degrees of freedom and tracing out the original ones.  The partition sum (\ref{Z}) can be written
\begin{eqnarray}
&& Z(K_{1},K_{2}) = \left( \cosh{K_{1}} \right)^{2N} \left( \cosh{K_{2}} \right)^{N} \sum_{\{s^{a},s^{b}\}} 
\nonumber \\
&& \prod_{\vt \in \Lambda^*} \left(1 + \tanh{K_1} s_{l_\vt}^a s_{r_\vt}^a s_{t_\vt}^a \right)
\left(1 + \tanh{K_1} s_{l_\vt}^b s_{r_\vt}^b s_{t_\vt}^b \right)
\nonumber \\
&&
\prod_{i \in \Lambda} \left(1 + \tanh{K_2} s_{i}^a s_{i}^b \right)
.
\nonumber
\end{eqnarray}
If we introduce auxiliary binary variables $\{\ts_\vt^a,\ts_\vt^b=0,1\}$ on the dual lattice, we can express the partition sum as 
\begin{eqnarray}
&& Z(K_{1},K_{2}) = \left( \cosh{K_{1}} \right)^{2N} \left( \cosh{K_{2}} \right)^{N} 
\sum_{\{s^{a},s^{b}\}} \sum_{\{\ts^{a},\ts^{b}\}} 
\nonumber \\
&& \prod_{\vt \in \Lambda^*} \left( \tanh{K_1} \right)^{\ts_\vt^a + \ts_\vt^b}
\left(s_{l_\vt}^a s_{r_\vt}^a s_{t_\vt}^a \right)^{\ts_\vt^a}
\left(s_{l_\vt}^b s_{r_\vt}^b s_{t_\vt}^b \right)^{\ts_\vt^b}
\nonumber \\
&&
\prod_{i \in \Lambda} \left(1 + \tanh{K_2} s_{i}^a s_{i}^b \right), 
\nonumber
\end{eqnarray}
and by rearranging the factors,
\begin{eqnarray}
&& Z(K_{1},K_{2}) = \left( \cosh{K_{1}} \right)^{2N} \left( \cosh{K_{2}} \right)^{N} 
\sum_{\{s^{a},s^{b}\}} \sum_{\{\s^{a},\s^{b}\}} 
\nonumber \\
&& \prod_{\vt \in \Lambda^*} \left( \tanh{K_1} \right)^{\frac{1}{2}(\s_\vt^a + \s_\vt^b) +1}
\prod_{i \in \Lambda} \left( s_{i}^a \right)^{\frac{1}{2}(\s_{l_i}^a \s_{r_i}^a \s_{b_i}^a + 1)} 
\nonumber \\
&&
\times \left( s_{i}^b \right)^{\frac{1}{2}(\s_{l_i}^b \s_{r_i}^b \s_{b_i}^b + 1)} 
\left(1 + \tanh{K_2} s_{i}^a s_{i}^b \right), 
\nonumber
\end{eqnarray}
where we have defined the auxiliary spins, $\s_\vt \equiv 2 \ts_\vt - 1 = \pm 1$.  In the above expression spins on different sites $i$ are uncoupled and can be traced out.  If we sum over the four possible values of the pair of spins $(s_i^a,s_i^b)$ the terms with $\s_{l_i}^a \s_{r_i}^a \s_{b_i}^a \neq \s_{l_i}^b \s_{r_i}^b \s_{b_i}^b$ vanish.  This means that all the auxiliary $a$ and $b$ plaquettes have to be same.  But for the chosen boundary conditions (see above) the spin-plaquette mapping is one-to-one, which means that $\s_\vt^a = \s_\vt^b$ for all $\vt$.  After summing over all the spins $(s^a,s^b)$ we therefore obtain,
\begin{eqnarray}
Z(K_{1},K_{2}) &=& 2^{N/2} \left( \sinh{2 K_{1}} \right)^{N} \left( \sinh{2 K_{2}} \right)^{N/2} 
\nonumber \\
&& \sum_{\{\s\}} 
\prod_{\vt \in \Lambda^*} \left( \tanh{K_1} \right)^{\s_\vt}
\prod_{i \in \Lambda} \left( \tanh{K_2} \right)^{\frac{1}{2} \s_{l_i} \s_{r_i} \s_{b_i}} 
,
\nonumber
\end{eqnarray}
where we have renamed $\s^a \to \s$.  Now we have a single set of spins $\s$, and the partition sum of the two coupled TPMs can be written as, 
\begin{equation}
Z(K_{1},K_{2}) = 2^{N/2} \left( \sinh{2 K_{1}} \right)^{N} \left( \sinh{2 K_{2}} \right)^{N/2} Y(\tK_1,\tK_2) ,
\label{ZZ}
\end{equation}
where $Y(\tK_1,\tK_2)$ is the partition sum of a {\em single} TPM in a field \cite{Sasa2010}, with 
\begin{equation}
e^{-2\tK_{1}} = \tanh{K_{2}}, \;\;\; e^{-\tK_2} = \tanh{K_{1}} .
\label{KK}
\end{equation}
From Ref.\ \cite{Sasa2010} we know that the TPM in a field has an exact duality, namely,
\begin{equation}
Y(\tK_{1},\tK_{2}) = \left( \sinh{2 \tK_{1}} \sinh{2 \tK_{2}} \right)^{N/2} Y(\tK_1^*,\tK_2^*) ,
\label{YY}
\end{equation}
with
\begin{equation}
e^{-2\tK_{1}^*} = \tanh{\tK_{2}}, \;\;\; e^{-2 \tK_2^*} = \tanh{\tK_{1}} .
\label{KKs}
\end{equation}
The partition sum of the dual TPM in a field, with paremeters $\tK_{1}^*,\tK_{2}^*$, in turn is equivalent to that of a pair of replicated TPMs, with parameters $K_{1}^*,K_{2}^*$, via Eqs.\ (\ref{ZZ})-(\ref{KKs}).  From this we recover the duality of the coupled TMPs given in Eqs.\ (\ref{Z})-(\ref{K}).

\noindent
{\bf Acknowledgements.} I am grateful to David Chandler for discussions.  This work was supported in part by Leverhulme Trust grant no.\ F/00114/BG.


%

\end{document}